\documentclass[10pt, pre, amsmath, onecolumn, showpacs, superscriptaddress]{revtex4-1}

\usepackage{amsmath}
\usepackage{amssymb}
\usepackage{mathtools}
\usepackage{esint}

\usepackage{color}
\definecolor{dgreen}{rgb}{0,0.7,0}

\def\bluew#1{{\color{blue} #1}}

\usepackage{caption}
\usepackage{subcaption}

\DeclareMathOperator{\sign}{sign}

\newcommand{\g}{g} %Single-particle propagator
\newcommand{\G}{\mathbb{G}} %Multi-particle propagator
\newcommand{\cM}{\mathcal{M}} %large deviation function of Y and theta
\newcommand{\cF}{\mathcal{F}} %large deviation function of Y only

\newcommand{\dd}{\mathrm{d}}
\newcommand{\ee}{\mathrm{e}}
\newcommand{\ci}{\mathrm{i}}
\newcommand{\cO}{\mathcal{O}}

\newcommand{\yl}{{y_\ell}} 
\newcommand{\bx}{\mathbf{x}}
\newcommand{\by}{\mathbf{y}}
\newcommand{\Sp}{\scriptstyle\mathfrak{S}}

\newcommand{\cQ}{\mathcal{Q}}
\newcommand{\cA}{\mathcal{A}}
\newcommand{\st}{{\sigma_t}}

\newcommand{\Lm}{L_-}
\newcommand{\Lp}{L_+}
\newcommand{\rhom}{\rho_-}
\newcommand{\rhop}{\rho_+}
\newcommand{\thetast}{{\theta^*}}

\newcommand{\rhoi}{\varrho}
\newcommand{\lambdat}{\tilde{\lambda}}

\begin{document}
\title{Tagged particle in single-file diffusion with arbitrary initial conditions}
\author{J. Cividini}
\address{Department of Physics of Complex Systems, Weizmann Institute of Science, Rehovot 76100, Israel}
\author{A. Kundu}
\address{International center for theoretical sciences, TIFR, Bangalore - 560012, India}
%\ead{anupam.kundu@icts.res.in}
\date{\today}

\begin{abstract}
%We consider a general single-file-problem with arbitrary initial position distributions of the particle for both the quenched and the annealed cases. 
We compute the full probability distribution of the positions of a tagged particle exactly for given arbitrary initial positions of the particles and for general single-particle propagators. We consider the thermodynamic limit of our exact expressions in quenched and annealed settings.  For a particular class of single-particle propagators, the exact formula is expressed in a simple integral form in the quenched case whereas in the annealed case, it is expressed as a simple combination of Bessel functions. In particular, we focus on the step and the power-law initial configurations. In the former case,  a drift is induced even when the one-particle propagators are symmetric. On the other hand, in the later case the scaling of the cumulants of the position of the tracer becomes different than the uniform case. We provide numerical verifications of our results.

\end{abstract}

\maketitle

\section{Introduction}

%\subsection{General intro}

%\greenw{We review the problem of tagged particles in single files and present some new results.}\\
\noindent
Determining the effective motion of a single particle in presence of other interacting particles is a well known fundamental problem in statistical physics. Such problems appear in various situations starting from 
motions of pollen grains in water, motion of colloidal particles in solutions, movements of molecules inside cell, 
traffic flows, inertial particle motions in turbulence to active particle motions like collective motions in flock of birds or school of fishes. When the time scales of different particles are well separated, one usually considers a simple phenomenological description of the system e.g. Langevin description. On the other hand if the time scales are comparable then such a phenomenology does not provide good descriptions and one needs to consider the full interacting many particle problem. In case of passive particles (the paper deals with only these kinds of particles), since the interaction among the particles are usually strongly repulsive at short distances and weakly attractive at large distances, any substantial displacement of one particle creates a density modulation in the vicinity of it, which in turn affects the motion of the surrounding particles. As a result the effective motion of a single particle becomes quite different than its free motion and also gets highly correlated with the motion of other particles. For example, a free single Brownian particle exhibits diffusive behaviour: its mean-square displacement grows linearly with time. However, the effective motion of the same Brownian particle in presence of other impenetrable diffusing particles gets changed: its `mobility' is reduced. In two and three dimensions, this effect may be significant in highly dense situations. But in one dimension this effect is the strongest, as the particles can not overtake each other and motion of one particle is hemmed by the presence of other particles. This makes the transport anomalously slow. The effective motion of a Brownian particle in the pool of other hard-core Brownian particle in one dimension becomes sub-diffusive, \emph{i.e.} the mean squared displacement grows as $\sim \sqrt{t}$ with time instead of increasing linearly as $~t$. Such motion of non-overtaking particles in one dimension or more realistically in narrow channels is known as single-file diffusion, where one is usually interested in the motion of a tagged particle (TP) which undergoes the same dynamics as the host particles. The only difference is that it just carries a mark.

Single file diffusion appears in various practical situations, for example, in ions passing through narrow pores in cell membrane \cite{Hodgkin55}, in transport in ionic conductors \cite{Richards77}, in protein sliding 
along DNA sequences \cite{Li09}, and also in motion of molecules inside a porous medium \cite{Krager92} or inside carbon nano-tubes \cite{Das10}. Theoretical study of such restricted motion of particles have been started long back by Harris \cite{Harris-65} and Jepsen \cite{Jepsen-65}. They showed that, the mean squared displacement (MSD) of a tagged particle grows diffusively when the individual particles evolve according to Hamiltonian dynamics whereas for Brownian particles the MSD of a TP grows subdiffusively as stated earlier. Since then there have been many theoretical analyses on single file diffusion
which, in addition to the anomalous behaviour, also suggests that the typical fluctuation of the displacement of the TP becomes Gaussian at late times \cite{Harris-65, Levitt73, Jara06, Arratia-83}.  Later in 1998 this fact was established in a more detailed calculation of the probability distribution of the TP displacement in single file systems with hardcore Brownian particles, which is also valid at finite times. While there are many results available for hard-core inter-particle interaction, very less theoretical results are available for more general interactions. These include: the simple exclusion process where the MSD and the fourth cumulant is known \cite{Krapivsky-14, Krapivsky-15}, and the colloidal systems where the MSD is known in terms of isothermal compressibility \cite{Kollmann-03}.

%Recently, several experiments have been able to observe TP diffusion by passive 
%microrehology in zeolites, transport of colloidal particles or charged spheres in narrow circular channels \cite{Gupta-95, Kulka-96, 
%Hahn-96, Wei-00, Meersmann-00, Lin-05, Lutz-04}.  
%Such experimental evidences have generated 
%a great revival of interest in tagged particle diffusion.
%Very recently many different results regarding the tagged particle diffusion have been reported for various systems with differently organized dynamics 
%\cite{Percus-74, Beijeren-83, Arratia-83, Alexander-78, Majumdar-91, Lizana-09, Barkai-09, Kollmann-03, Gupta-07, Rodenbeck-98,
% Barkai-10, Roy-13, Sabhapandit-07, Illien-13, Benichou-13, Krapivsky-14}. 
%For example, in addition to the mean position and the MSD of the TP, the large deviation functions (LDF) associated to the probability distribution functions (PDF) of the displacement of the TP have been studied as well \cite{Barkai-09, Barkai-10, Lizana-09, 
%Krapivsky-14, Hegde-14, Illien-13, Sabhapandit-15}. 
%%\greenw{Elaborate on the MFT derivation and microscopic derivation of LDF}. 

Several experiments have recently been able to observe TP diffusion, for example, in passive 
microrehology in zeolites, transport of colloidal particles or charged spheres in narrow circular channels \cite{Gupta-95, Kulka-96, 
Hahn-96, Wei-00, Meersmann-00, Lin-05, Lutz-04}.  
Such experimental evidences have re-generated a great interest in tagged particle diffusion. Consequently, many different results on mean position, the MSD and also the large deviation functions (LDF) associated to the probability distribution functions (PDF) of the displacement of the TP \cite{Barkai-09, Barkai-10, Lizana-09, 
Krapivsky-14, Hegde-14, Illien-13, Sabhapandit-15}, 
have very recently been reported for several systems with different multiparticle dynamics 
\cite{Percus-74, Beijeren-83, Arratia-83, Alexander-78, Majumdar-91, Lizana-09, Barkai-09, Kollmann-03, Gupta-07, Rodenbeck-98,
 Barkai-10, Roy-13, Sabhapandit-07, Illien-13, Benichou-13, Krapivsky-14}.  

In all these studies it was noted that the MSD as well as the form of the LDF depend on the choice of the initial configurations (IC) in the thermodynamic limit. Until now, two types of initial states have mainly been studied which we call uniform quenched and uniform annealed ICs. In the uniform annealed case, the initial position of the particles are drawn randomly from a distribution with average uniform density $\rho_0$.  On the other hand, in the uniform quenched case, initially particles are placed at uniform separation $\rho_0^{-1}$, where $\rho_0$ is the density. In the context of  Random Average Process, Rajesh and Majumdar have first shown that the $\sqrt{t}$ scaling of the MSD is same but the pre-factors are different in the two settings \cite{Rajesh-01}. Later a similar result was also obtained in the context of hard-core Brownian particles \cite{Barkai-10}. In the last few years it has also been shown that the forms of the LDF associated to the distribution of the displacement of the TP are different in the above mentioned two cases : uniform quenched and uniform annealed ICs. For hard-core Brownian particles, these LDFs are derived using a hydrodynamic description called the macroscopic fluctuation theory \cite{Krapivsky-14, Krapivsky-15} as well as from the microscopic point of view using a mapping to non-interacting Brownian particles \cite{Sadhu-15}. Another microscopic derivation of the LDF considers general individual single particle propagators but only for the uniform annealed ICs \cite{Hegde-14}. 

In this paper, we provide microscopic derivations of the \emph{full} distribution of the TP 
displacement for \emph{arbitrary} initial configurations and general propagators. This allows us to compute the distribution for arbitrary inhomogeneous initial density profiles in the thermodynamic limit. In particular, we provide explicit exact expressions of this distribution for the step initial configuration as well as for power law initial configurations (explained later), both in the quenched and the annealed case. In addition to providing alternate expressions of the distributions, we also show that our results are consistent with the existing results in the uniform/flat case.  Especially, for power law ICs we show that not only the form of the LDF changes but also the scaling with respect to time changes.

We consider $2N+1$ identical hard-core particles moving on the real line where each individual particle is described by a free propagator $ \g(y,t|x,0)$. The quantity $ \g(y,t|x,0)$ represents the transition probability that the particle reaches the position $y$ at time $t$, starting from $x$.
%\begin{equation}
% \label{single-pro}
% \g(y,t|x,0)=\frac{1}{\sigma_t} G\left( \frac{y-x}{\st} \right),
%\end{equation}
%with a characteristic time-dependent length scale $\st$ (see~\cite{Hegde-14, Sabhapandit-15}) where $\st$ is a monotonically increasing function of $t$. The quantity $ \g(y,t|x,0)$ represents the transition probability the particle reaches the position $y$ at time $t$, starting from $x$. Normalization $\int_{y=-\infty}^\infty  \g(y,t|x,0) \dd y = 1$ implies $\int_{u=-\infty}^\infty  G(u) \dd u = 1$ and we take $G(u)$ to be symmetric $G(u)=G(-u)$.  One example of this form of propagators is the brownian propagator with $G(u) = \exp \left( - {u^2}/{2} \right)/\sqrt{2 \pi} $ and $\st=\sqrt{2t}$. Another example would be for Hamiltonian dynamics where initial velocities chosen independently from Gaussian distribution with zero mean and variance $\bar{v}^2$. In this case one has a Gaussian propagator with $\st=\bar{v}^2t$. 
Because of the hard-core interaction among the particles, they can not cross each other and thus keep their initial order maintained. Hence the many particle propagator $\G_{2 N+1}(\by,t|\bx,0)$ representing the transition probability that the $2N+1$ particles, starting from positions $\bx=(x_{-N},x_{-N+1},..,x_0,...x_{N-1},x_N)$ with $x_i \le x_{i+1}$ reach the positions $\by=(y_{-N},y_{-N+1},..,y_0,...y_{N-1},y_N)$ with $y_i \le y_{i+1}$ at time $t$, is given explicitly by \cite{Rodenbeck-98}
\begin{equation}
	 \label{eq:gnyx}
	\G_{2 N+1}(\by,t|\bx,0) = \sum_{\tau \in \Sp_{2N+1}} \prod_{k=-N}^N \g(y_k,t|x_{\tau(k)},0),
\end{equation}
where the sum is over the set $\Sp_{2N+1}$ of all possible permutations of $2N+1$ distinct elements.  By definition the $(2N+1)$-particle propagator is normalised, $\int_{-\infty < y_{-N} < \ldots < y_N < \infty} \G_{2 N+1}(\by,t|\bx,0) \dd \by = 1$. 
Initial positions of the particles are chosen within the ranges
$-L_-\leq x_{-N}< x_{-N+1}<..<x_0<...<x_{N-1}<x_N\leq L_+$ and in the computation, we will take the thermodynamic limit $N \to \infty$ and $L_{\pm} \to \infty$ keeping the density finite. These initial configurations (ICs) are usually chosen in the following two ways:
\begin{itemize}
\item \emph{Quenched}: The initial positions of the particles at $t=0$ are exactly fixed. The superscript $\cQ$ will be used for this setup.
\item \emph{Annealed}: The initial positions of the particles are drawn randomly from a distribution. The superscript $\cA$ will be used for this setup.
\end{itemize}
Note that the final positions $\by$ can be anywhere on the infinite line. The multi-particle propagator~\eqref{eq:gnyx} contains all the information about the system. Since at each collision between two identical and interacting hard-point particles one particle is acting as a reflecting hard wall for the other particle, one can effectively treat the whole interacting system as non-interacting by exchanging their identities \cite{Jepsen-65, Hegde-14, Sadhu-15}. Hence, in the non-interacting picture one can think that the particles just pass through each other when they collide and each particle executes an independent motion which is described by the propagator in eq.\eqref{single-pro}.

We are interested in the distribution $P(y_0,t|\bx)$ of the displacement made by the tagged particle ($0$th particle) in time $t$ for given initial positions $\bx$ of all the particles. This is done by first performing the ordered $...<y_i<y_{i+1}<...$ integrals over $y_i$s except $y_0$ and then taking the thermodynamic limit for both\bluew{,} quenched and annealed case.

The paper is organised as follows: In sec.~\ref{section:galcalc} we derive an exact expression \eqref{eq:Pyt1} of the TP propagator $P(y_0,t|\bx,0)\equiv P(y_0,t|\bx)$ for arbitrary initial positions of the other particles. Starting from this expression we, in the next section, take the thermodynamic limit for arbitrary initial density profiles in both the quenched and the annealed setup separately. We then focus on particular cases in secs. \ref{section:step} and \ref{section:ldfgal} where we consider the step and the 
power-law ICs, respectively. In the step case, the initial densities are different on the right and left of the tagged particle whereas in the power law ICs the initial density are symmetric but inhomogeneous. In both cases, exact formulas are obtained for the full distribution of the tagged particle displacements, from which one can easily get the corresponding LDFs. This generalises previously known results for the LDF in the uniform initial density profile case~\cite{Hegde-14, Sabhapandit-15, Sadhu-15}. In particular for power-law initial density profiles, we show that the  scaling of the cumulants with respect to time becomes different than the same in the uniform case. In section~\ref{section:ccl} we conclude the paper.

%\greenw{Say sth about prefactors?}. 
%In the next section, by taking same densities on both sides in the results from sec. \ref{section:step}, we reproduce the LDF associated to the distribution of TP displacements for uniform initial density fields.  Finally, in section~\ref{section:ldfgal} we find the form of the large deviation function for arbitrary initial density profiles again both in the quenched and annealed case. 

\section{TP propagator for arbitrary initial positions of the particles}
\label{section:galcalc}
\noindent
To derive the TP propagator $P(y_0,t|\bx)$  for arbitrary initial positions of the particles 
$(x_{-N},x_{-N+1},..,x_{-1},x_0,x-1,...x_{N-1},x_N)$. We start with the multi-particle propagator in \eqref{eq:gnyx} and integrate over all the final positions except the TP final position :
\begin{equation}
\label{eq:defPyt}
%P(y,t) = \int_{-\infty < y_{-N} < \ldots < y_{-1} < y} \int_{y < y_1 < \ldots < y_N < \infty} \sum_{\sigma \in \Sp_{2N+1}}
%\prod_{k=-N}^N g(y_k,t|x_{\sigma(k)},0) \dd \by_{\backslash 0},
P(y_0,t|\bx) = \int_{-\infty}^{y_{1-N}}dy_{-N}  \int_{-\infty}^{y_{2-N}}dy_{1-N}...\int_{-\infty}^{y_0}dy_{-1}  \int_{y_0}^{y_2}dy_{1}  \int_{y_0}^{y_3}dy_{2}... \int_{y_0}^{\infty}dy_{N}
\sum_{\sigma \in \Sp_{2N+1}}
\prod_{k=-N}^N g(y_k,t|x_{\sigma(k)},0) ,
\end{equation}
%Equation~\eqref{eq:defPyt} can be simplified to~\eqref{eq:Pyt} for any intiial condition $\bx$. Remember that, TP initially is always at $x_0=0$. 
We first decompose the sum over permutations in $\Sp_{2N+1}$ into two parts. In the first part we sum over the choices of \emph{unordered} sets $S^- = \{\sigma(-N), \ldots, \sigma(-1)\}$, $\{\sigma(0)\}$ and $S^+ = \{\sigma(1), \ldots, \sigma(N)\}$ and in the second part, for each such choices we sum over the permutations $\tau^-$ and $\tau^+$ of $N$ elements in sets $S^-$ and $S^+$, respectively. As a result we have
\begin{eqnarray}
\label{eq:Pyt}
P(y_0,t|\bx) &=& \sum_{S^+, m, S^-} g(y_0,t|x_{m},0) \sum_{\tau^- \in \Sp_{N}} \int_{-\infty}^{y_{1-N}}dy_{-N}  \int_{-\infty}^{y_{2-N}}dy_{1-N}...\int_{-\infty}^{y_0}dy_{-1} \prod_{k=-N}^{-1} g(y_k,t|x_{S^-[\tau^-(k)]},0), \nonumber \\ 
&&~~~~~ \times \sum_{\tau^+ \in \Sp_{N}}  \int_{y_0}^{y_2}dy_{1}  \int_{y_0}^{y_3}dy_{2}... \int_{y_0}^{\infty}dy_{N} 
\prod_{k=-1}^{N} g(y_k,t|x_{S^+[\tau^+(k)]},0). 
\end{eqnarray}
Clearly the sets $S^-$, $\{\sigma(0)\}$ and $S^+$ contain the starting positions of those particles that have arrived at positions $y_k<y_0$, $y_0$ and $y_k>y_0$, respectively, while the permutations $\tau^+$ and $\tau^-$ represent their precise order. 
Now using the following facts 
\begin{eqnarray}
\sum_{\tau^+ \in \Sp_{N}}  \int_{y_0}^{y_2}dy_{1}  \int_{y_0}^{y_3}dy_{2}... \int_{y_0}^{\infty}dy_{N} 
\prod_{k=1}^{N} g(y_k,t|x_{S^+[\tau^+(k)]},0) &=& 
\prod_{k=1}^{N} \int_{y_0}^{\infty} g(y_k,t|x_{S^+[k]},0) \dd y_k, \nonumber \\
\sum_{\tau^- \in \Sp_{N}} \int_{-\infty}^{y_{1-N}}dy_{-N}  \int_{-\infty}^{y_{2-N}}dy_{1-N}...\int_{-\infty}^{y_0}dy_{-1} \prod_{k=-N}^{-1} g(y_k,t|x_{S^-[\tau^-(k)]},0) &=& 
\prod_{k=-1}^{-N} \int_{\infty}^{y_0} g(y_k,t|x_{S^-[k]},0) \dd y_k, 
\end{eqnarray}
in \eqref{eq:Pyt} and defining the functions 
$g_+(x;y,t) = \int_{y}^\infty  \dd y'~g(y',t|x,0)$ and $g_-(x;y,t) = \int_{-\infty}^y  \dd y'~g(y',t|x,0)$, 
%\begin{eqnarray}
%g_+(x;y,t) = \int_{y'=y}^\infty g(y',t|x,0) \dd y', ~~~~~~ 
%%\nonumber \\ 
%g_-(x;y,t) = \int_{y'=-\infty}^y g(y',t|x,0) \dd y',
%\end{eqnarray}
we have 
\begin{eqnarray}
\label{eq:Pyt0}
P(y_0,t|\bx) &=&\sum_{m=-N}^Ng(y_0,t|x_{m},0)  \sum_{\{S^+, S^-\}} \prod_{\stackrel{k=-N}{k\neq m}}^{-1} g_-(x_{S^-[k]};y_0,t) \prod_{\stackrel{k=1}{k\neq m}}^{N} g_+(x_{S^+[k]};y_0,t),
\end{eqnarray}
where we are left with the sums over choices of the sets $\{S_+, x_m,S_-\}$. The above expression has the following interpretation in the non-interacting picture: the term $g(y_0,t|x_{m},0)$ is present due to the fact that,  the particle with initial position $x_m$ (\emph{i.e.} the initial $m$th particle) now has become the middle particle \emph{i.e.} $0$th particle at time $t$. Given this fact, among the rest of the $2N$ particles 
with initial positions $(x_{-N},...,x_{m-1},x_{m+1},...,x_N)$, we choose $N$ of them which go to the left of $y_0$ and the remaining $N$ go to the right of $y_0$ at time $t$. Thus both the sets $S_+$ and $S_-$ contains exactly $N$ elements and the sum over $\{S^+,S_-\}$ is exactly over these choices. Each of the terms of this sum looks like $\prod_{k=-N,\neq m}^N g_{\epsilon_k}(x_k;y_0,t)$ with $\epsilon_k=\pm1$ and there are exactly $N$ terms whose $\epsilon$ values are $-1$ and the remaining terms have $\epsilon=1$. Note that, $g_{1}(x_k;y_0,t)$ and $g_{-1}(x_k;y_0,t)$ represent the probability of finding the particle at positions greater than and less than $y_0$ at time $t$ given that it had started at position $x_k$. Introducing the delta function $\delta_{\sum_{k=-N,k\neq m}^{N}\epsilon_k,0} $ which ensures that there are exactly $N$ particles in each set,  we can write \cite{Sadhu-15},  
\begin{eqnarray}
\label{eq:Pyt1}
P(y_0,t|\bx) &=& \sum_{m=-N}^N g(y_0,t|x_{m},0) \prod_{k=-N,k\neq m}^N \left(\sum_{\epsilon_{k}=\pm}\right) \delta_{\sum_{k=-N,k\neq m}^{N}\epsilon_k,0} 
\prod_{k=-N, k\neq m}^{N} g_{\epsilon_{k}}(x_k;y_0,t)\label{discrete_p-y} \nonumber \\
&=& \frac{1}{2 \pi} \int_{-\pi}^{\pi} \dd \theta\prod_{k=-N}^{N} (\ee^{\ci \frac{\theta}{2}} g_+(x_k;y_0,t)+\ee^{-\ci \frac{\theta}{2}} g_-(x_k;y_0,t)) 
\sum_{m=-N}^N \frac{g(y_0,t|x_m,0)}{\ee^{\ci \frac{\theta}{2}} g_+(x_m;y_0,t)+\ee^{-\ci \frac{\theta}{2}} g_-(x_m;y_0,t)}, \nonumber \\
&=& \frac{\dd}{\dd y_0} \left[ -\frac{1}{4 \pi \ci} \int_{-\pi}^\pi  \frac{\dd \theta }{\sin \left( {\theta}/{2} \right)} \prod_{k=-N}^{N} (\ee^{\ci \frac{\theta}{2}} g_+(x_k;y_0,t)+\ee^{-\ci \frac{\theta}{2}} g_-(x_k;y_0,t)) \right].
\end{eqnarray}
where we have used $\delta_{n,0}=(1/2\pi) \int_{-\pi}^\pi e^{in\theta}d\theta$.
The last line of the above equation is our first result.  Starting from this equation one can easily check that this distribution is normalised as follows : Integrating both sides of \eqref{eq:Pyt1}, we have 
\begin{eqnarray}
\int_{-\infty}^\infty dy_0~P(y_0,t|\bx) = 
\frac{1}{2 \pi} \int_{-\pi}^\pi  \dd \theta ~\frac{ \sin \left( (2N+1)\frac{\theta}{2} \right)}{\sin \left( \frac{\theta}{2} \right)} = 1. \label{norm-chk-1}
\end{eqnarray}
%In the next section, we will take thermodynamic limit \emph{i.e.} take $N \to \infty$ and $L_\pm \to \infty$ in this expressions keeping the density finite for both the quenched and annealed settings. 
%In the quenched case, 

%\newpage
\section{Thermodynamic limit}
\label{Thermodynamic-limit}

\noindent
In the previous section we have obtained an explicit expression for the TP displacement distribution in 
\eqref{eq:Pyt1} for given arbitrary initial positions of the $(2N+1)$ particles. In this section we will take thermodynamic limit \emph{i.e.} $N \to \infty$ and $L_\pm \to \infty$ limit of \eqref{eq:Pyt1} for both the quenched and annealed case while keeping the densities finite.  From now onwards we will consider the following general class of the propagators 
\begin{equation}
 \label{single-pro}
 \g(y,t|x,0)=\frac{1}{\sigma_t} G\left( \frac{y-x}{\st} \right),
\end{equation}
with a characteristic time-dependent length scale $\st$ (see~\cite{Hegde-14, Sabhapandit-15}) where $\st$ is a monotonically increasing function of $t$. In addition we assume $G(u)$ to be symmetric $G(u)=G(-u)$. Normalisation $\int_{y=-\infty}^\infty  \g(y,t|x,0) \dd y = 1$ implies $\int_{u=-\infty}^\infty  G(u) \dd u = 1$.  One example of this form of propagators is the brownian propagator with $G(u) = \exp \left( - {u^2}/{2} \right)/\sqrt{2 \pi} $ and $\st=\sqrt{2t}$. Another example would be for Hamiltonian dynamics where initial velocities chosen independently from Gaussian distribution with zero mean and variance $\bar{v}^2$. In this case also one has a Gaussian propagator with mean zero and variance $\st=\bar{v}^2t$. 
%\greenw{Actually I don't think the mapping works in that case: particles with exclusion should exchange their velocities whereas independent particles don't.}

 Let us start with the quenched case.

\subsection{Quenched case}

\noindent
In this case particles are started at positions $x_k$ given by some well defined functions 
$x_k=f_-(k,L_-,N)$ for $k = -N, \ldots, -1$, $x_k=f_+(k,L_-,N)$ for $k = 1,\ldots, N$ and $x_0=0$. 
For example, in step ICs, $f_\pm(k, L_\pm, N)= k\frac{2 L_\pm}{2 N+1}$. 
Let us first look at the thermodynamic limit of the product term inside the $\theta$ integral : 
\begin{eqnarray}
\label{eq:prodlim}
 \prod_{k=-N}^{N} (\ee^{\ci \frac{\theta}{2}} g_+(x_k;y_0,t)+\ee^{-\ci \frac{\theta}{2}} g_-(x_k;y_0,t))  &=& (1- g_+(x_0;y_0,t)(1-\ee^{\ci \theta}))^{1/2} \prod_{k=-N}^{-1} (1- g_+(x_k;y_0,t)(1-\ee^{\ci \theta})) \nonumber \\ 
 && \times (1- g_-(x_0;y_0,t)(1-\ee^{-\ci \theta}))^{1/2}
 \prod_{k=1}^{N} (1- g_-(x_k;y_0,t)(1-\ee^{-\ci \theta})). 
% &\stackrel{=}{N \rightarrow \infty}& \exp\left[ \sum_{k=-\infty}^{-1} \log \left( 1-g_+(x_k;y,t) (1-\ee^{\ci \theta})\right)+ \frac{1}{2} \log \left( 1-g_+(x_0;y,t) (1-\ee^{\ci \theta})\right) \right. \nonumber \\ && \left.  + \frac{1}{2} \log \left( 1-g_-(x_0;y,t) (1-\ee^{-\ci \theta})\right) + \sum_{k=1}^{\infty} \log \left( 1-g_-(x_k;y,t) (1-\ee^{-\ci \theta})\right) \right], \nonumber 
\end{eqnarray}
Taking the thermodynamic limit in the above equation we have
%When time is large $\st$ becomes large as well and the sums can be approximated by integrals. 
%We get
\begin{eqnarray}
\label{eq:prodlimtime}
RHS&=&\exp\left[ \sum_{k=-\infty}^{-1} \log \left( 1-g_+(x_k;y_0,t) (1-\ee^{\ci \theta})\right)+ \frac{1}{2} \log \left( 1-g_+(x_0;y_0,t) (1-\ee^{\ci \theta})\right) \right. \nonumber \\ 
&& \left.  + \frac{1}{2} \log \left( 1-g_-(x_0;y_0,t) (1-\ee^{-\ci \theta})\right) + \sum_{k=1}^{\infty} \log \left( 1-g_-(x_k;y_0,t) (1-\ee^{-\ci \theta})\right) \right]  \nonumber \\
&=&\exp\left[ \int_{-\infty}^0 dx~ \rhoi_-(x)  \log \left( 1-g_+(x;y_0,t) (1-\ee^{\ci \theta})\right)+ 
\int_0^{\infty} dx~ \rhoi_+(x)  \log \left( 1-g_-(x;y_0,t) (1-\ee^{-\ci \theta})\right) 
 \right] \nonumber 
%&=& \exp\left[ \st \cM^\cQ_{\rhom,\rhop}(Y,\theta) + \cO(\st^{-1})\right], \nonumber
\end{eqnarray}
where $\rhoi_+(x)=\sum_{k=1}^{\infty}\delta(x-f_+(k,L_+,N)) + \delta(x)/2$ and $\rhoi_-(x)=\sum_{k=-\infty}^{-1}\delta(x-f_-(k,L_-,N)) + \delta(x)/2$. The scaling form of $g(y,t|x,0)$ in \eqref{single-pro}, implies that $g_\pm(x;y,t)$ also has the following form $g_\pm(x;y,t) = G_\pm\left(\frac{y-x}{\st}\right)$, where $G_+(X) = \int_{X}^\infty G(Z) \dd Z$ and $G_-(X) = \int_{-\infty}^X G(Z) \dd Z$.
Using these scaling functions in \eqref{eq:Pyt1} and performing some manipulations  
we obtain
\begin{eqnarray}
%  &=& \frac{1}{2 \pi} \int_{\theta = -\pi}^\pi  \frac{\dd \theta }{2 \ci \sin \left( \frac{\theta}{2} \right)} \left(\rhom \log\left(1+(\ee^{\ci \theta}-1) G_+(Y)\right) + \rhop \log\left(1+(\ee^{- \ci \theta}-1) G_-(Y)\right) \right) \ee^{\st \cM^\cQ_{\rhom,\rhop}(Y,\theta)} \nonumber \\
 P^\cQ_{\rhoi} (y_0,t)&=& \frac{\dd}{\dd y_0} \left[ -\frac{1}{4 \pi \ci} \int_{-\pi}^\pi  \frac{\dd \theta }{\sin \left( \frac{\theta}{2} \right)} \exp \left({\st~\cM^\cQ_t\left(\frac{y_0}{\st},\theta\right)}\right) \right],~~{\text{up~to}~\cO(\st^{-1})},~~\text{where}, \label{P-Q} \\
 \cM^\cQ_t(Y,\theta) &=& \int_{-\infty}^\infty \dd Z \left[ \Theta(-Z) \rhoi_-(\st Z) \log \left( 1+ (\ee^{\ci \theta}-1) G_+(Y-Z) \right) \right. \nonumber \\
&& ~~~~~~~~~~~~~~~~~~~~~~~~~~~~~~~~~~~~~~~~
\left.+ \Theta(Z) \rhoi_+(\st Z) \log \left( 1+ (\ee^{-\ci \theta}-1) G_-(Y-Z) \right) \right] ,
~~~~~ \label{cM-Q}
\end{eqnarray}
with $\Theta(Z)$ as the Heaviside theta function. Note that the quantity within brackets in \eqref{P-Q} is the cumulative distribution of $y_0$. One can easily check that this distribution is normalised as follows : Integrating both sides of \eqref{P-Q}, we have 
\begin{eqnarray}
\int_{-\infty}^\infty dy_0~P^\cQ_{\rhoi} (y_0,t) = 
-\frac{1}{4 \pi \ci} \int_{-\pi}^\pi  \frac{\dd \theta }{\sin \left( \frac{\theta}{2} \right)}  \left[ \exp \left({\st~\cM^\cQ_t\left(\frac{y_0}{\st},\theta\right)}\right) \Bigg{|}_{y\to \infty} - \exp \left({\st~\cM^\cQ_t\left(\frac{y_0}{\st},\theta\right)}\right) \Bigg{|}_{y\to -\infty} \right]. \nonumber 
\end{eqnarray}
Using the definitions of $G_\pm(Z)$ given before \eqref{cM-Q}, one can easily see that 
$\st \cM^\cQ_{\rhom,\rhop}(y_0/\st,\theta)_{y \to \pm \infty} \to \mp \ci \theta N$. As a result we have 
\begin{eqnarray}
\int_{-\infty}^\infty dy_0~P^\cQ_{\rhoi} (y_0,t) &=& 
-\frac{1}{4 \pi \ci} \int_{-\pi}^\pi  \frac{\dd \theta }{\sin \left( \frac{\theta}{2} \right)}  \left[ \exp (-\ci \theta N) - \exp(\ci \theta N) \right]_{N \to \infty}, \nonumber \\
&=& \frac{1}{2 \pi }  \int_{-\pi}^\pi \dd \theta \left(  \frac{\sin( N \theta) }{\sin ( \theta/2)} \right)_{N \to \infty} = \frac{1}{\pi }  \int_{-\infty}^\infty \dd \psi~\frac{\sin(\psi) }{\psi} = 1.
\end{eqnarray}
The calculation presented here gives an exact expression of the full probability distribution of the TP displacement for arbitrary initial density profiles in the quenched setting. 

To get the expressions for the uniform annealed case, one has to use $\rhoi_\pm(x)=\rho$ in \eqref{P-Q} and \eqref{cM-Q}. As a result, the expression of the probability distribution $P^{Uni,\cQ}_{\rho} (y_0,t)$ becomes simpler which after some simplification reads as
\begin{equation}
\label{eq:Pflatq}
P^{Uni,\cQ}_{\rho} (y_0,t) = \frac{\rho}{2\pi} \int_{\theta=-\pi}^{\pi} \dd \theta \frac{\theta}{2 \sin \left( \frac{\theta}{2}\right)} 
\ee^{\st \cM^{Uni,\cQ}_{\rho} \left( Y,\theta \right)}
\end{equation}
where $\cM^{Uni,\cQ}_{\rho} \left( Y,\theta \right)$ given in~\eqref{cM-Q}.
Many results have been obtained for this case using various methods \cite{Krapivsky-15, Sadhu-15}. 
One can show that our results indeed reduce to those previously known results as follows : 
Let us start with the exponential term inside the $\theta$ integral in \eqref{P-Q}:
%Introducing $E(Y) = (1 - 2 G_+(Y))^2 -1 = (2 G_-(Y) - 1)^2 - 1$ in \eqref{cM-Q-step} we have,
\begin{eqnarray}
\label{eq:Mthetasimp}
\rho^{-1} \cM^{Uni,\cQ}_{\rho}(Y,\theta) &=& \int_{U=Y}^\infty \log \left( 1+ (\ee^{\ci \theta}-1) G_+(U) \right)\dd U + \int_{U=-\infty}^Y \log \left( 1+ (\ee^{-\ci \theta}-1) G_-(U) \right)\dd U \allowdisplaybreaks  \nonumber \\
&=& \int_{U=Y}^\infty \log\left[\ee^{\ci \frac{\theta}{2}} \left(\ee^{-\ci \frac{\theta}{2}}+2 \ci \sin \left( \frac{\theta}{2} \right) G_+(U)\right)\right] 
\dd U + \int_{U=-\infty}^Y \log\left[\ee^{-\ci \frac{\theta}{2}} \left(\ee^{-\ci \frac{\theta}{2}}+2 \ci \sin \left( \frac{\theta}{2} \right) G_+(U)\right)\right] \dd U \allowdisplaybreaks \nonumber \\
&=& - \ci \theta Y + \int_{U=-\infty}^\infty \log\left[\ee^{-\ci \frac{\theta}{2}}
+2 \ci \sin \left( \frac{\theta}{2} \right) G_+(U)\right] \dd U \allowdisplaybreaks \\
&=& -\ci \theta Y +\frac{1}{2} \int_{U=-\infty}^\infty \log\left[1+\sin \left( \frac{\theta}{2} \right)^2 ((2 G_+(U)-1)^2-1)\right] \dd U 
\allowdisplaybreaks \nonumber \\ 
&&~~~~~~~~~~~~~~~~~~~~~~~~~~~~~~~~+ \ci \int_{U=-\infty}^\infty \arctan \left[ \tan \left( \frac{\theta}{2} \right) (2 G_+(U)-1)\right] \dd U \allowdisplaybreaks \nonumber 
\end{eqnarray} 
Since $G(U)$ is an even function the last integral vanishes by symmetry. This gives the very simple result
\begin{eqnarray}
\label{eq:Mthetasimpf}
\cM^\cQ_{\rho}(Y,\theta) &=& \frac{\rho}{2}\int_{U=-\infty}^\infty \log \left( 1+ 4 \sin^2 \left( \frac{\theta}{2} \right) G_+(U) G_-(U) \right) \dd U - \ci \rho \theta Y,
\end{eqnarray} 
where we see that the real part is completely independent of $Y$, while the imaginary part does not depend on the process considered and has a very simple form. 
The result is the same as the one obtained in~\cite{Krapivsky-15, Sadhu-15}, which is expressed as a function of $B = \ci \theta$.  From equation~\eqref{eq:Mthetasimpf} the large deviation function can be obtained in an implicit form : $\cF^{Uni,\cQ}_{\rho}(Y)= \min_{\theta} \left[\cM^{Uni,\cQ}_{\rho}(Y,\theta) \right]$.
%\begin{equation}
%\cF^{Uni,\cQ}_{\rho,\rho}(Y)= \min_{\theta} \left[\cM^{Uni,\cQ}_{\rho,\rho}(Y,\theta) \right]
%\end{equation}

\subsection{Annealed case}

\noindent
In the annealed case, one chooses the initial positions of the $(2N+1)$ particles randomly from a distribution.  In this paper we first put the tagged particle at the origin \emph{i.e.} $x_0 = 0$. Next we randomly draw the positions of $N$ particles between $-\Lm$ and $0$ on the left of TP ($0$th particle) and the positions of the remaining $N$ particles are drawn between $0$ and $\Lp$ on the right of TP such that the average initial density becomes $\rhoi_-(x)$ and $\rhoi_+(x)$ on left and right side respectively.  In fact this is obtained by placing $N$ particles randomly on the negative half-axis with probability density $\frac{\rhoi_-(x)}{N}$ and placing the remaining $N$ particles are placed on the positive half-axis with probability density $\frac{\rhoi_+(x)}{N}$. After choosing $N$ positions, we label them according to their order on both sides. For example, on the right side the particles labelled as $k=1, \ldots, N$ have positions $0<...<x_k<x_{k+1}...<L_+$. Similarly on left side we label the particles as  $-1, \ldots, -N$.  

For each realisation of the initial positions $\bx = (x_{-N},...,x_{-1},0,x_1,...,x_N)$, 
we have the propagator $P(y_0,t|\bx)$.  Averaging over initial positions, the probability of the TP displacement, can be expressed as 
\begin{eqnarray}
P^\cA_{\rhoi}(y_0,t) &=& N! \int_{-L_-}^0 dx_{-1} \frac{\rhoi_-(x_{-1})}{N} 
 \int_{-L_-}^{x_{-1}} dx_{-2} \frac{\rhoi_-(x_{-2})}{N} ...\int_{-L_-}^{x_{-N+1}} dx_{-N} \frac{\rhoi_-(x_{-N})}{N}   
 \nonumber \\ 
&&~~~~~~~~~~ \times 
 N! \int_{0}^{L_+} dx_{1} \frac{\rhoi_+(x_{1})}{N}  
 \int_{x_1}^{L_+} dx_{2} \frac{\rhoi_+(x_{2})}{N} ...\int_{x_{N-1}}^{L_+} dx_{N} \frac{\rhoi_+(x_{N})}{N}  ~P(y_0,t|\bx), \nonumber \\
  &=&\frac{\dd}{\dd y_0} \left[ -\frac{1}{4 \pi \ci} \int_{-\pi}^\pi  \dd \theta~ \frac{ 1}{\sin \left( {\theta}/{2} \right)}  \left ( \int_{-L_-}^{0} dx \frac{\rhoi_-(x)}{N} (\ee^{\ci \frac{\theta}{2}} g_+(x;y_0,t)+\ee^{-\ci \frac{\theta}{2}} g_-(x;y_0,t)) \right)^N \right. \nonumber \\ 
&& \times (\ee^{\ci \frac{\theta}{2}} g_+(0;y_0,t)+\ee^{-\ci \frac{\theta}{2}} g_-(0;y_0,t)) \times \left. \left(  \int_0^{L_+} dx \frac{\rhoi_+(x)}{N} (\ee^{\ci \frac{\theta}{2}} g_+(x;y_0,t)+\ee^{-\ci \frac{\theta}{2}} g_-(x;y_0,t)) \right )^N+ \frac{1}{2} \right] \nonumber \\
 &=&\frac{\dd}{\dd y_0} \left[ \frac{1}{2} -\frac{1}{4 \pi \ci} \int_{-\pi}^\pi  \dd \theta~ \left( \frac{\cos \left( {\theta}/{2}\right)}{\sin \left( {\theta}/{2}\right)} + \ci (g_+(0;y_0,t) - g_-(0;y_0,t))\right) 
 %\frac{( \cos \left( {\theta}/{2} \right) + (2g_+(0;y_0,t)-1) \ci \sin \left( {\theta}/{2} \right))}{\sin \left( {\theta}/{2} \right)} 
 \right. \nonumber \\ 
&\times& \left.\left ( \int_{-L_-}^{0} dx \frac{\rhoi_-(x)}{N} (\ee^{\ci \frac{\theta}{2}} g_+(x;y_0,t)+\ee^{-\ci \frac{\theta}{2}} g_-(x;y_0,t)) \right)^N   \left(  \int_0^{L_+} dx \frac{\rhoi_+(x)}{N} (\ee^{\ci \frac{\theta}{2}} g_+(x;y_0,t)+\ee^{-\ci \frac{\theta}{2}} g_-(x;y_0,t)) \right )^N\right]. \nonumber
\end{eqnarray}
Taking the $N \to \infty$ and $L_\pm \to \infty$ limit while keeping the density finite, one has 
\begin{eqnarray}
 P^\cA_{\rhoi}(y,t) &=&\frac{\dd}{\dd y_0} \left[ -\frac{1}{4 \pi \ci} \int_{-\pi}^\pi  \dd \theta~\left( \frac{\cos \left( {\theta}/{2}\right)}{\sin \left( {\theta}/{2}\right)} + \ci (g_+(0;y_0,t) - g_-(0;y_0,t))\right) 
 %\frac{( \cos \left( {\theta}/{2} \right) + (2g_+(0;y_0,t)-1) \ci \sin \left( {\theta}/{2} \right))}{\sin \left( {\theta}/{2} \right)} 
 \exp \left(\st \cM^\cA_{\rhoi}\left(\frac{y_0}{\st},\theta \right) \right) \right ],~~{\text{up~to}~\cO(\st^{-1})},  \label{P-A} \\
\text{where}, && \cM^\cA_{\rhoi}(Y,\theta) = (\ee^{\ci \theta}-1) \int_{-\infty}^0 \dd Z ~G_+(Y-Z) \rhoi_-(\st Z)   +  (\ee^{-\ci \theta}-1)\int_0^{\infty} \dd Z ~G_-(Y-Z) \rhoi_+(\st Z). \label{cM-A}
%\cM^\cA_{\rhoi}(Y,\theta) &=& (\ee^{\ci \theta}-1) \int_{Y}^\infty \dd U G(U) \int_{Y-U}^0 \dd Z \rhoi(\st Z)   +  (\ee^{-\ci \theta}-1) \int_{-\infty}^Y \dd U G(U) \int_{0}^{Y-U} \dd Z \rhoi(\st Z), 
\label{eq:defMA}
\end{eqnarray}
These two equations together provide the distribution of the TP displacements in annealed setup with arbitrary initial average density. To check the normalisation we follow the same as done in the quenched case in the previous section. Integrating both sides of \eqref{P-A} with respect to $y_0$ and performing some manipulations, we have 
\begin{eqnarray}
\int_{-\infty}^\infty dy_0~P^\cA_{\rhoi} (y_0,t) 
%&=&\frac{\ee^{-N}}{\pi } \int_{-\pi}^\pi  \dd \theta~\exp \left(N\ee^{-\ci \theta}\right) 
% +\frac{\ee^{-N}}{2 \pi \ci} \int_{-\pi}^\pi  \dd \theta \frac{ \cos \left(\theta/ 2\right)}{\sin \left(\theta/2 \right)}~
% ~\exp \left(N\ee^{\ci \theta}\right) \\
&=& \frac{\ee^{-N}}{2\pi } \int_{-\pi}^\pi  \dd \theta \left[ 1
 -\ci\frac{ \cos \left(\theta/ 2\right)}{\sin \left(\theta/2 \right)} \right]
 ~\exp \left(N\ee^{\ci \theta}\right)=  \frac{\ee^{-N}}{\pi \ci } \ointctrclockwise dz~\frac{\ee^{Nz}}{z-1}=1
  \label{norm-A-1}
\end{eqnarray}
%\begin{eqnarray}
%\int_{-\infty}^\infty dy_0~P^\cA_{\rhoi} (y_0,t) &=& 
%-\frac{1}{4 \pi \ci} \int_{-\pi}^\pi  \dd \theta~\frac{ \cos \left(\theta/ 2\right)}{\sin \left(\theta/2 \right)}  \left[ \exp \left({\st~\cM^\cA_t\left(\frac{y_0}{\st},\theta\right)}\right) \Bigg{|}_{y\to \infty} - \exp \left({\st~\cM^\cA_t\left(\frac{y_0}{\st},\theta\right)}\right) \Bigg{|}_{y\to -\infty} \right], \allowdisplaybreaks \nonumber \\
%&=&-\frac{1}{4 \pi \ci} \int_{-\pi}^\pi \dd \theta ~\frac{ \cos \left(\theta/ 2\right)}{\sin \left(\theta/2 \right)}
% \left[ \exp \left(N(\ee^{-\ci \theta}-1)\right)  - \exp \left(N(\ee^{-\ci \theta}-1)\right) \right]_{N \to \infty},
%\end{eqnarray}
where we have used the definitions of $G_\pm(Z)$ given before \eqref{cM-Q} and
$\int_0^\infty dx~\rhoi_+(x) = \int_{-\infty}^0 dx~\rhoi_-(x)=N$. 
In the last part of \eqref{norm-A-1} we used the transformation $z=\ee^{\ci \theta}$. As a result, the integral over $\theta$ in\eqref{norm-A-1} gets converted to an integral on the complex plane along an unit circle (centred around the origin) counterclockwise. This integral on the complex plane can easily be performed using contour integral technique to find the value one. 
%\begin{eqnarray}
%\int_{-\infty}^\infty dy_0~P^\cA_{\rhoi} (y_0,t) 
%&=&-\frac{1}{2 \pi } \int_{-\pi}^\pi  \dd \theta~\frac{ \cos \left(\theta/ 2\right)}{\sin \left(\theta/2 \right)}~
% \left[ \sin \left( N \sin \theta\right)~\ee^{-N(1-\cos \theta)} \right]_{N \to \infty},\nonumber \\
% &=& \frac{1}{2 \pi }  \int_{-\pi}^\pi \dd \theta \left(  \frac{\sin( N \theta) }{\sin ( \theta/2)} \right)_{N \to \infty} = \frac{1}{\pi }  \int_{-\infty}^\infty \dd \psi~\frac{\sin(\psi) }{\psi} = 1.
%\end{eqnarray}
%In the last line of the above equation array we have used the fact that in the $N \to \infty$ the weight of the exponential term is highly concentrated over small theta \emph{i.e.} around $\theta \sim 0$. 

Furthermore, one can perform the $\theta$ integral in \eqref{P-A} using modified Bessel functions $I_n(q) =  \frac{1}{2 \pi} \int_{-\pi}^\pi  \dd \theta \ee^{q\cos \theta} \ee^{\pm \ci n \theta}$ as follows:  
%One can express the $\theta$ integral in terms of Bessel functions. 
Let us define $A(y_0,t)$ and $B(y_0,t)$ as
\begin{equation}
 \label{defAB}
 A(y_0,t) = \st  \int_{-\infty}^0 \dd Z ~G_+(Y-Z) \rhoi_-(\st Z),~~B(y_0,t)=\st \int_0^{\infty} \dd Z ~G_-(Y-Z) \rhoi_+(\st Z),
\end{equation}
such that $\st \cM^\cA_{\rhoi}(Y,\theta) = (\cos \theta -1) (A+B) + \ci \sin \theta (A-B)$. In the second term [proportional to $(g_+-g_-)$] in \eqref{P-A}, we can use the identity 
\begin{equation}
 \frac{1}{2 \pi} \int_{\theta=-\pi}^\pi \ee^{(A+B) \cos \theta + \ci (A-B) \sin \theta} \dd \theta  = I_0(2 \sqrt{AB}). \nonumber
\end{equation}
%where $I_n(q) =  \frac{1}{2 \pi} \int_{-\pi}^\pi  \dd \theta \ee^{q\cos \theta} \ee^{\ci n \theta}$ is a modified Bessel function of the first kind. Then we can take the $y_0$ derivative using $\frac{\dd I_0(q)}{\dd q} = I_1(q)$.
%$I_n(q) =  \frac{1}{2 \pi} \int_{-\pi}^\pi  \dd \theta \ee^{q\cos \theta} \ee^{\ci n \theta}$.
Now to simplify the term proportional to $\frac{\cos \left( {\theta}/{2}\right)}{\sin \left( {\theta}/{2}\right)}$ one can first take the $y_0$ derivative, which produces $(\cos \theta -1)$ and $\sin \theta$ factors. These can be simplified further and eventually be expressed as derivatives of $\exp \left[\st \cM^\cA_{\rhoi}(Y,\theta)\right] = \exp\left[ (\cos \theta -1) (A+B) + \ci \sin \theta (A-B) \right]$ with respect to $A$ and $B$, which can again be expressed as a function of $I_0$.
After some simplifications, we finally find
\begin{eqnarray}
P^\cA_{\rhoi}(y_0,t) = \frac{\ee^{-A-B}}{\st} &&\left[ \left[ G +\st \left(G_+ B' - G_- A' \right)\right] I_0 \left( 2 \sqrt{AB}\right)+ \st \left( G_- \sqrt{\frac{A}{B}} B' - G_+\sqrt{\frac{B}{A}} A' \right) I_1 \left( 2 \sqrt{AB}\right)\right],
\label{P-A-simp}
\end{eqnarray}
where the $A$, $B$ functions are evaluated at $(y_0,t)$ and the $G$, $G_\pm$ functions at $\frac{y_0}{\st}$.  The derivatives of  $A$ and $B$ with respect to $y_0$ are denoted by $A'$ and $B'$, respectively and we used
$I_1(q)={\dd I_0(q)}/{\dd q}$.
%
%Furthermore, using the following identity for the modified Bessel function $I_0(q) =  \frac{1}{2 \pi} \int_{-\pi}^\pi  \dd \theta \ee^{q\cos \theta}$
%%\begin{equation}
%%I_0(q)= \frac{1}{2 \pi} \int_{-\pi}^\pi  \dd \theta \ee^{q\cos \theta}, \nonumber
%%\end{equation}
%one can perform the $\theta$ integral completely in \eqref{P-A} to get 
%% as follows : \redw{I will add the details later.} Finally we get 
%\begin{eqnarray}
%P^\cA_{\rhoi}(y_0,t) &=& \frac{e^{-C(y_0,t)}}{2}\left[ \left(\frac{d D}{d y_0}\right)\left(\frac{\partial }{\partial C}+1\right) + \left(\frac{d C}{d %y_0}\right)\left(\frac{\partial }{\partial D}\right)\right]~I_0\left(\sqrt{C(y_0,t)^2-D(y_0,t)^2}\right) \nonumber \\ 
%&& -\frac{1}{2}\frac{d}{dy_0} \left[ (G_+(y_0/\st)-G_-(y_0/\st))~e^{-C(y_0,t)}I_0\left(\sqrt{C(y_0,t)^2-D(y_0,t)^2}\right) \right] ,\nonumber \\
%&=& \frac{e^{-C(y_0,t)}}{2} \left \{ \left[ \left(\frac{d D}{d y_0}\right)\left(\frac{\partial }{\partial C}+1\right) + \left(\frac{d C}{d y_0}\right)\left(\frac{\partial }{\partial D}+(G_+(y_0/\st)-G_-(y_0/\st))\right)\right]~I_0\left(\sqrt{C(y_0,t)^2-D(y_0,t)^2}\right) \right. \nonumber \\ 
%&& -\left. \frac{d}{dy_0} \left[ (G_+(y_0/\st)-G_-(y_0/\st))~I_0\left(\sqrt{C(y_0,t)^2-D(y_0,t)^2}\right) \right] \right \},~~~\text{where},\nonumber \\
% && C(y_0,t) = A(y_0,t) +B(y_0,t),~~~~D(y_0,t) = A(y_0,t) - B(y_0,t),~~\text{and}, \nonumber \\
%A(y_0,t) &=& \st  \int_{-\infty}^0 \dd Z ~G_+(Y-Z) \rhoi_-(\st Z),~~B(y_0,t)=\st \int_0^{\infty} \dd Z ~G_-(Y-Z) \rhoi_+(\st Z). ~~\redw{\text{need~to~improve}}\nonumber 
%\end{eqnarray}
The expression in \eqref{P-A-simp} provides an exact and explicit expression for the distribution of the displacement made by the TP in the annealed case for arbitrary initial average density. 
As long as the initial density profile is such that $2 \sqrt{A B} \to \infty$ for large times, one can use the asymptotic form $I_n(q) \sim \frac{\ee^{q}}{\sqrt{2 \pi q}} (1+\cO(q^{-1}))$ for large $q$. Hence, for large times in 
the annealed case the large deviation function is essentially given  by 
\begin{eqnarray}
 \label{F-A-any}
% \log P^\cA_{\rhoi}(y_0,t) \simeq -A(y_0,t)-B(y_0,t) + 2 \sqrt{A(y_0,t) B(y_0,t)} = - \left[ \sqrt{A(y_0,t)}-\sqrt{B(y_0,t)}\right]^2,
  \log P^\cA_{\rhoi}(y_0,t) &\simeq & - \left[ \sqrt{A(y_0,t)}-\sqrt{B(y_0,t)}\right]^2,
%  = - \st \left[ \sqrt{\int_{-\infty}^0 \dd Z ~G_+(Y-Z) \rhoi_-(\st Z)}-\sqrt{\int_0^{\infty} \dd Z ~G_-(Y-Z) \rhoi_+(\st Z)}\right]^2, 
\end{eqnarray}
where the functions $A(y_0,t)$ and $B(y_0,t)$ are given in \eqref{defAB}.

For the uniform annealed case where the average densities are $\rhoi_\pm(x)=\rho$,  the large deviation function associated to this probability density function for Brownian particles has been obtained earlier using macroscopic fluctuation theory~\cite{Krapivsky-15, Sadhu-15}. For general propagators of the form \eqref{single-pro}, the full probability distribution has also been derived in~\cite{Hegde-14, Sabhapandit-15} for this case. One can easily check that our expression for $P^\cA_{\rhoi}(y_0,t)$ in \eqref{P-A-simp} for arbitrary initial average density, reduces to these previous results for uniform annealed case. To do that, one should use the following simplified expressions: $A(y_0,t) = \st \rho \left( Q(Y) - \frac{Y}{2}\right)$, $B(y_0,t) = \st \rho \left( Q(Y) + \frac{Y}{2}\right)$ with $Q(Y)$ as in~\cite{Hegde-14,Sabhapandit-15},
\begin{equation}
 Q(Y) = Y \int_{Z=0}^Y G(Z) \dd Z +\int_{Z=Y}^\infty Z G(Z) \dd Z,
 \label{defQ}
\end{equation}
and $\frac{\dd A}{\dd y_0} = -\rho G_+(y_0/\st)$ and $\frac{\dd B}{\dd y_0} = \rho G_-(y_0/\st)$ for $\rhoi_\pm(x)=\rho$.
%With these one can easily check that our above expression for $P^\cA_{\rhoi}(y_0,t)$ reduces to these previous results for equal densities. 

\section{Step initial condition}
\label{section:step}
\noindent
Here we consider the step initial condition, where the starting densities are constant but different on the right and left of the tagged particle. We fix the initial position of the TP at $x_0=0$ and impose two different densities on each side in both the quenched and annealed case. Even though the single-particle propagator $G(U)$ is symmetric in space,  the density difference, as we will see, induces a drift of the TP. We are interested in computing this 
drift as well as the distribution of TP displacement $y_0$. Let us again start with the quenched 
case.

\subsection{Quenched case}

\begin{figure}[h]
  \centering
      \includegraphics[width=0.8\textwidth]{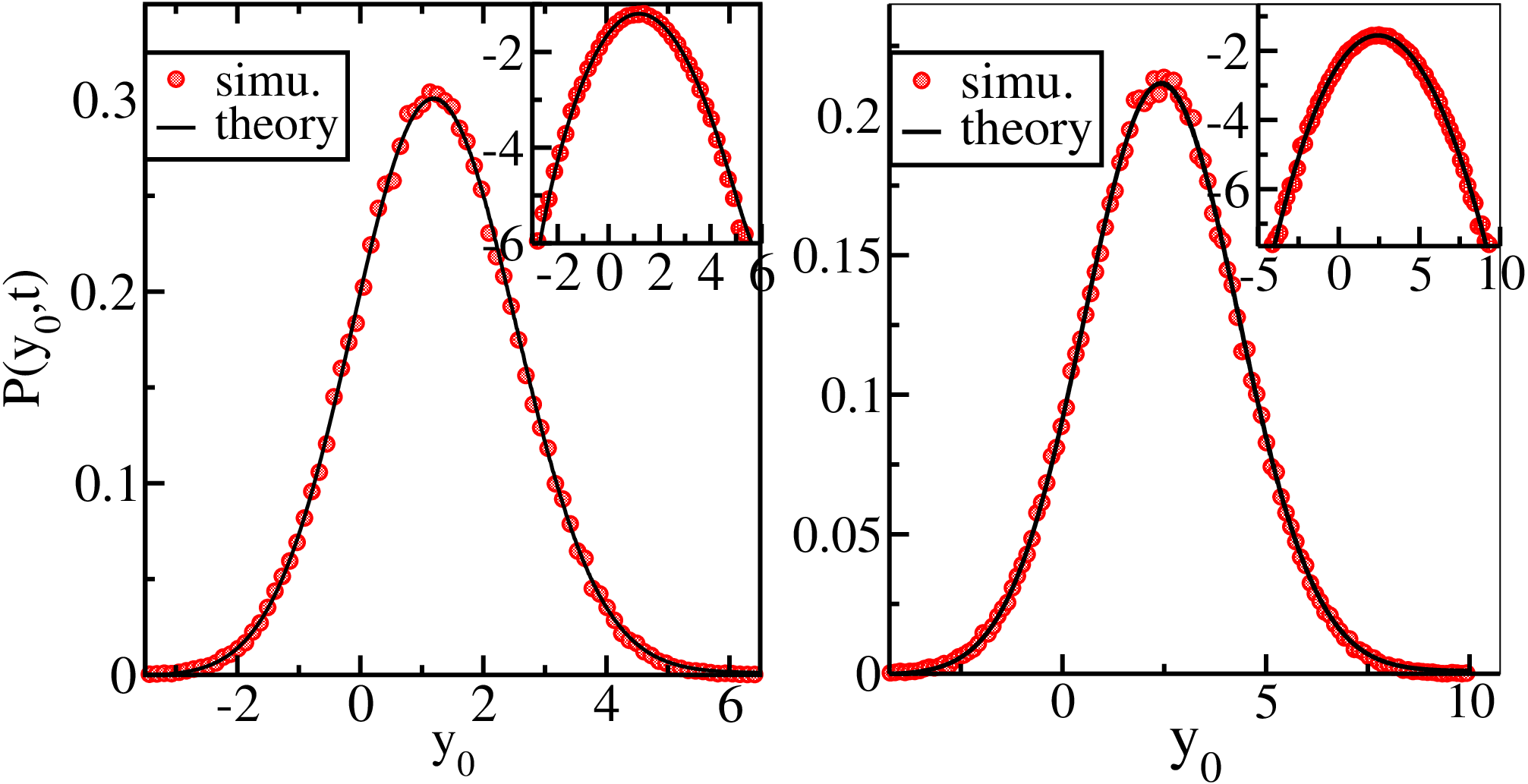}
  \caption{Tagged particle position probability in the quenched case with a step initial condition for densities $\rhom=2$ and $\rhop=1$ for two different times $t=10$ (left) and $t=40$ (right). For each figure the main plot is in linear scale while in the inset the vertical scale  is logarithmic. The particles are Brownian: $\st = \sqrt{2 t}$, $G(X) = \frac{\exp\left({-X^2/2}\right)}{\sqrt{2 \pi}}$.
%   $G_{\pm}(X) = \frac{1 \mp \Erf\left({X}/{\sqrt{2}}\right)}{2}$.
}
\label{fig:Psq}
\end{figure}

\noindent
The initial positions of the particles, in this case are $x_k = k\frac{2 \Lm}{2 N+1}$ for $k = -N, \ldots, -1$, $x_k = k\frac{2 \Lp}{2 N+1}$ for $k = 1,\ldots, N$ and $x_0 = 0$. We are interested in the $N \to \infty$ and $L_\pm \to \infty$ limit of \eqref{eq:Pyt1} while keeping $\rhom = \frac{N}{\Lm}$ and $\rhop = \frac{N}{\Lp}$ fixed. As a result the density profile is $\rhoi(x) = \rho_+\Theta(x) + \rho_-\Theta(-x)$. Using this density in \eqref{P-Q} and \eqref{cM-Q} we have 
\begin{eqnarray}
 \label{eq:Pstepquenched}
%  &=& \frac{1}{2 \pi} \int_{\theta = -\pi}^\pi  \frac{\dd \theta }{2 \ci \sin \left( \frac{\theta}{2} \right)} \left(\rhom \log\left(1+(\ee^{\ci \theta}-1) G_+(Y)\right) + \rhop \log\left(1+(\ee^{- \ci \theta}-1) G_-(Y)\right) \right) \ee^{\st \cM^\cQ_{\rhom,\rhop}(Y,\theta)} \nonumber \\
 P^{Step,\cQ}_{\rhom,\rhop} (\st Y,t)&=& \frac{\dd}{\dd y_0} \left[ -\frac{1}{4 \pi \ci} \int_{\theta = -\pi}^\pi  \frac{\dd \theta }{\sin \left( \frac{\theta}{2} \right)} \exp\left({\st~\cM^{Step,\cQ}_{\rhom,\rhop}\left(\frac{y_0}{\st},\theta \right)}\right) \right],~~
 \text{where}, \\
 \cM^{Step,\cQ}_{\rhom,\rhop}(Y,\theta)&=& \rhom \int_{Z=-\infty}^0 \log \left( 1+ (\ee^{\ci \theta}-1) G_+(Y-Z) \right)\dd Z + \rhop \int_{Z=0}^\infty \log \left( 1+ (\ee^{-\ci \theta}-1) G_-(Y-Z) \right)\dd Z. \label{cM-Q-step}
\end{eqnarray}
From these expressions one can easily compute the large deviation function through Legendre transform.  For large times, the $\theta$ integral could in principle be carried out by saddle-point approximation, however, the solution can be written only in an implicit way which is neither simple nor illuminating. However, it is evident from \eqref{cM-Q-step} that the LDF for this case matches with the same as obtained in \cite{Krapivsky-14, Krapivsky-15}. For smaller times the integrals in \eqref{eq:Pstepquenched} can be evaluated numerically. In fig.~\ref{fig:Psq} we 
compare the numerical evaluation of the theoretical expression of $P^{Step,\cQ}_{\rhom,\rhop} (\st Y,t)$ for Brownian particles with the same obtained by simulating the microscopic dynamics for two times $t=10$ and $t=40$. 
In both cases, we observe nice agreements even for quite small values of the probabilities (of the order of $\sim 10^{-6}$ for $t=10$ and $\sim 10^{-8}$ for $t=40$).

\subsection{Annealed case}
\begin{figure}[h]
  \centering
      \includegraphics[width=0.8\textwidth]{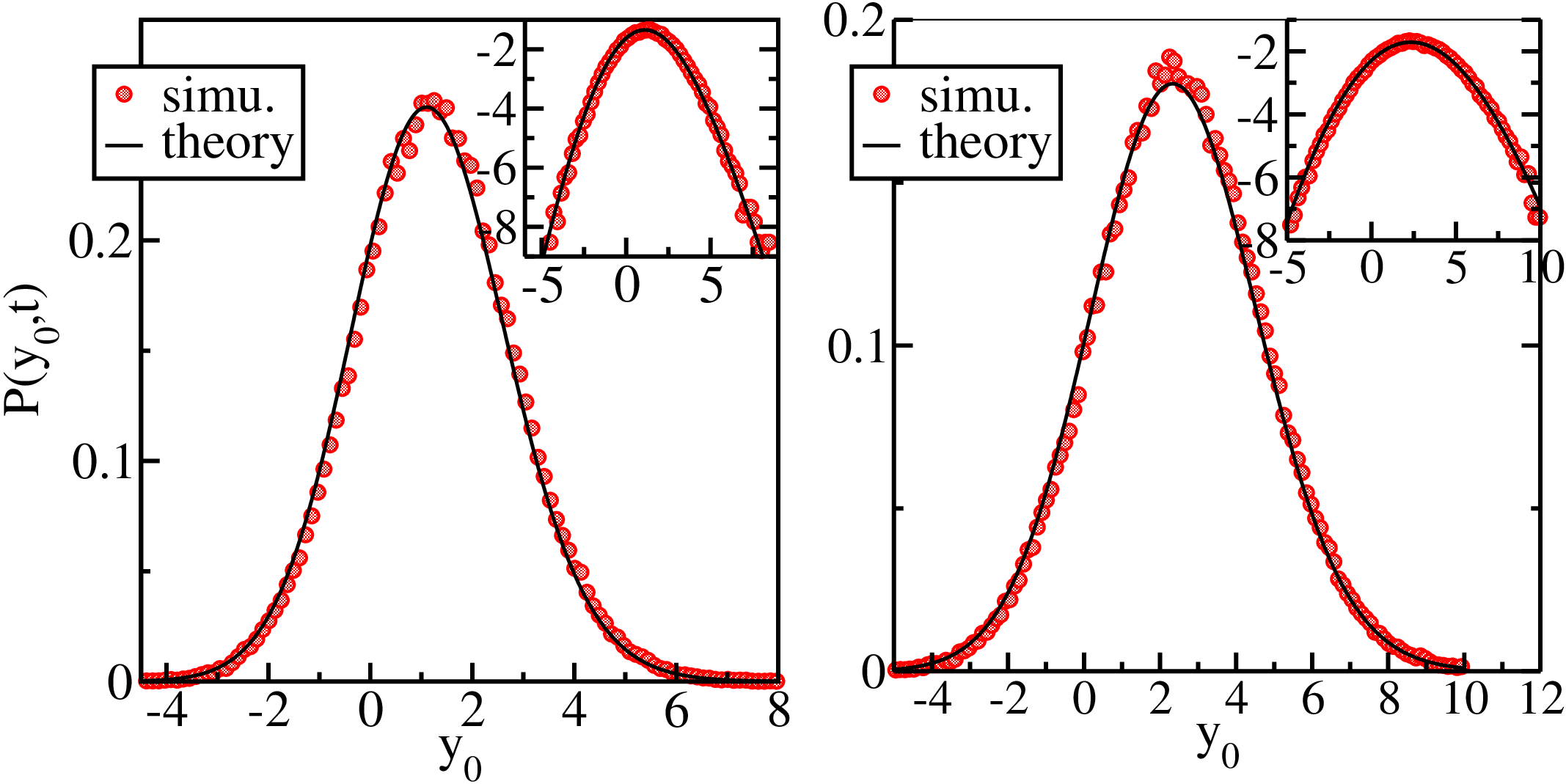}
  \caption{Tagged particle position probability in the annealed case with a step initial condition for densities $\rhom=2$ and $\rhop=1$ for two different times $t=10$ (left) and $t=40$ (right). For each figure the main plot is in linear scale while the vertical scale in the inset is logarithmic. The particles are Brownian: $\st = \sqrt{2 t}$, $G(X) = \frac{\exp\left({-X^2/2}\right)}{\sqrt{2 \pi}}$.
%   $G_{\pm}(X) = \frac{1 \mp \Erf\left(\frac{X}{\sqrt{2}}\right)}{2}$.
}
\label{fig:Psa}
  \end{figure}

\noindent
In this case, keeping $x_0 = 0$, we draw the positions of $N$ particles $-N, \ldots, -1$ uniformly between $-\Lm$ and $0$ and the positions of $N$ other particles $1, \ldots, N$ between $0$ and $\Lp$ in such a way that the average density, on the left becomes $\rho_-$ and on the right becomes $\rho_+$. This implies $\rhoi_\pm(x)=\rho_\pm$, inserting which in \eqref{defAB} we obtain $A(y_0,t) = \st \rhom \left( Q(Y) - \frac{Y}{2}\right)$, $B(y_0,t) = \st \rhop \left( Q(Y) + \frac{Y}{2}\right)$, $A'(y_0,t) = -\rhom G_+(Y)$ and $B'(y_0,t) = \rhop G_-(Y)$. Using these expressions in \eqref{P-A-simp} yields
\begin{eqnarray}
 P^{Step,\cA}_{\rhom,\rhop}(y_0,t) &=& \frac{\ee^{-\st \left( (\rhop+\rhom) Q +\frac{\rhop-\rhom}{2} Y\right)}}{\st} 
 \left[ \left( G +\st (\rhop + \rhom) G_+ G_- \right) I_0 \left( \st \sqrt{\rhop \rhom} \sqrt{4 Q^2-Y^2}\right) \right. 
\label{P-A-step} \\ 
&&~~~~~~~~~~~~~~~~~~~~~
\left.+ \st \left( \rhop G_-^2 \sqrt{\frac{2Q-Y}{2Q+Y}} + \rhom G_+^2 \sqrt{\frac{2Q+Y}{2Q-Y}} \right) I_1 \left( \st \sqrt{\rhop \rhom} \sqrt{4 Q^2-Y^2}\right)\right]. \nonumber 
\end{eqnarray}
For Brownian particles, the distribution~\eqref{P-A-step} is compared to simulations for two different times ($t=10$ and $t=40$) in figure~\ref{fig:Psa}. Like in the quenched case, the agreement is very good even for very small probabilities. 

In the very large time limit, one can get a bit simpler expression for $P^{Step,\cA}_{\rhom,\rhop}(y_0,t)$ with $\rhoi_\pm(x)=\rho_\pm$, by performing the $\theta$ integral in \eqref{P-A} using the saddle point method. 
We look for a saddle point $\thetast$ such that $\partial_\theta \cM^\cA_{\rhom,\rhop}|_\thetast = 0$. This gives
\begin{equation}
 \label{eq:thetast}
 \ee^{2 \ci \thetast} = \frac{\rhop \left(Q(Y) + \frac{Y}{2} \right)}{\rhom \left( Q(Y) -\frac{Y}{2}\right)},
\end{equation}
using which we, after some simplifications, get 
\begin{eqnarray}
 P^{Step,\cA}_{\rhom,\rhop}(y_0,t) &=& \frac{(\rhop + \rhom) G_+ G_- \sqrt{4 Q^2-Y^2}+  \rhop G_-^2 (2Q-Y) + \rhom G_+^2 (2Q+Y)}{\sqrt{2 \pi \st} (\rhop \rhom)^{1/4} (4 Q^2 -Y^2)^{3/4}} 
~ \ee^{-\st \cF^{Step,\cA}_{\rhom,\rhop}(Y) } ~~\label{P-A-step-lt}
 \text{where}, \\
 \cF^{Step,\cA}_{\rhom,\rhop}(Y)&=&-\cM^\cA_{\rhom,\rhop}(Y,\thetast)= \left( \sqrt{\rhop \left( Q+\frac{Y}{2}\right)} - \sqrt{\rhom \left( Q-\frac{Y}{2}\right)} \right)^2. \label{cM-A-step}
\end{eqnarray}
The same result can be obtained by using the known large-argument behaviour of the Bessel functions.

%\section{Uniform initial condition}
%\label{section:uniform}
%
%\noindent
%In this section we present the results for initial uniform density in both the quenched and annealed case.
%Specific results for the uniform density case can be obtained by putting $\rhop = \rhom=\rho$ in the expressions for a step initial condition. We show that the expressions given in section~\ref{section:step} are consistent with existing literature for annealed case. In addition we present some new results for the quenched case.
%
%\subsection{Quenched case}
%\noindent
%
%
%\subsection{Annealed case}

\section{Power law initial condition}
\label{section:ldfgal}

\noindent
We consider an initial density profile different from the uniform and step cases. For both the quenched and annealed settings, we here consider that the density of particles  decreases as a power of the distance measured from the position of the initial middle particle or the TP. Interestingly in this case, as we will see, the scaling with respect to time is different from $\st$ but depends continuously on this power. 

\subsection{Quenched case}
\begin{figure}
  \centering
      \includegraphics[width=0.75\textwidth]{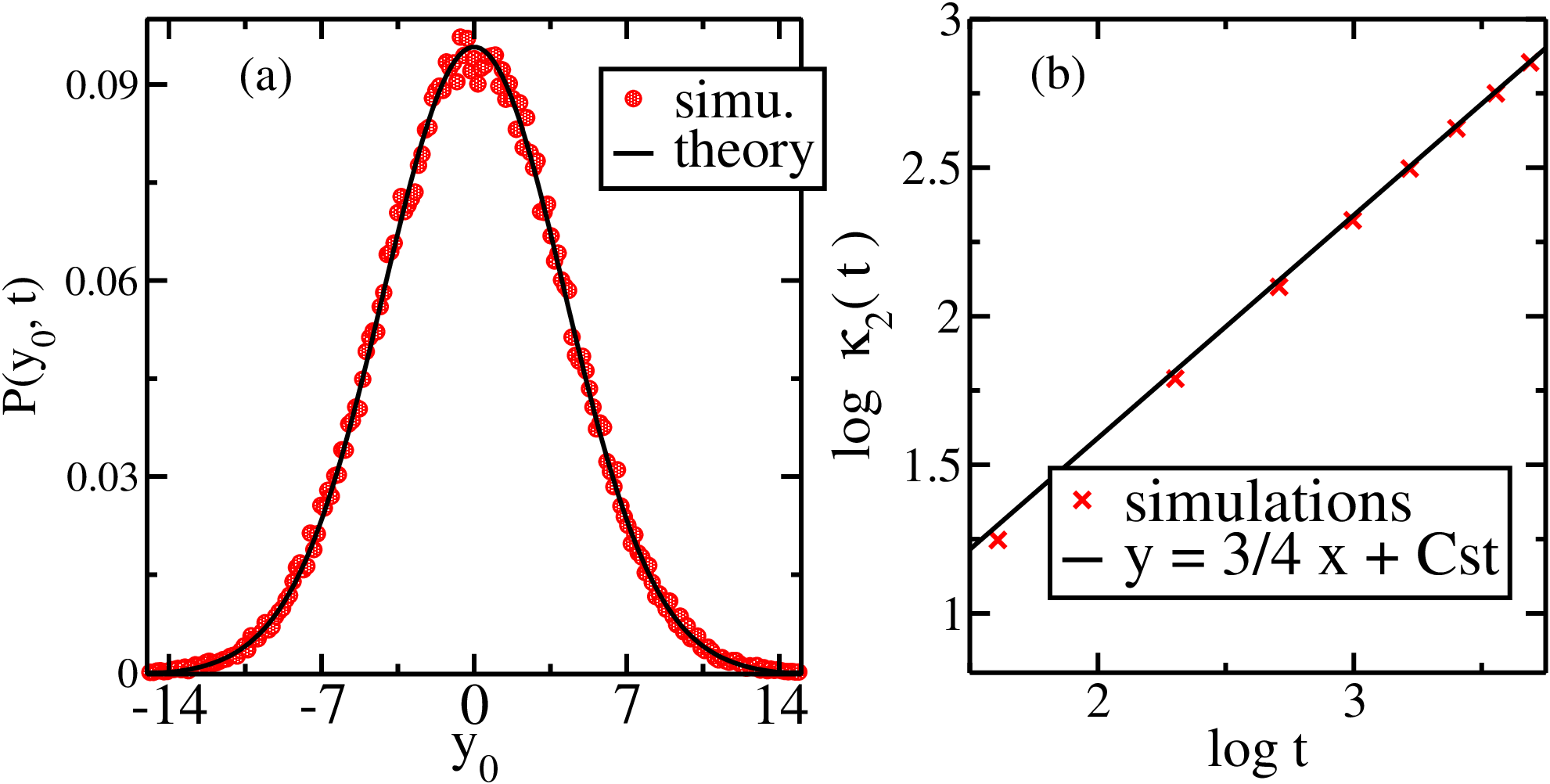}
  \caption{(a) Log-log plot of the variance $\kappa_2$ of the tagged particle position as a function of time and 
 (b) the tagged particle position probability at time $t=40$ for quadratic starting positions $\alpha=2$, $A=1$ and brownian particles $\st = \sqrt{2 t}$. The theoretical straight line has slope $3/4$. }
\label{fig:varCIC}
  \end{figure}

\noindent
In the quenched case,  power law initial density profiles can be obtained by placing the particle on both sides of the middle particle symmetrically but inhomogeneously. In particular we start with the following configuration  $x_k = A \sign(k) |k|^\alpha$, $\alpha > 0$ of the initial positions of the particles. This arrangement of the particles corresponds, in the thermodynamic limit, to the following power law density :
\begin{equation}
 \label{eq:rhoipow}
 \rhoi(x=\st Z) = \frac{2}{A (|k+1|^\alpha-|k-1|^\alpha)} \simeq \frac{1}{A \alpha} |k|^{1-\alpha} = \frac{1 }{A^{\frac{1}{\alpha}} \alpha} \st^{\frac{1}{\alpha}-1} |Z|^{\frac{1}{\alpha}-1},~~~\text{where},~~Z=\frac{|x|}{\st}.
\end{equation}
Using this form of $\rhoi(Z)$ in \eqref{P-Q} and \eqref{cM-Q} we get 
\begin{eqnarray}
 P^{PL,\cQ}_{\rhoi} (y_0,t)&=& \frac{\dd}{\dd y_0} \left[ -\frac{1}{4 \pi \ci} \int_{-\pi}^\pi  \frac{\dd \theta }{\sin \left( \frac{\theta}{2} \right)} \exp \left({\st^{1/\alpha}~\cM^{PL,\cQ}_\rhoi\left(\frac{y_0}{\st},\theta\right)}\right)  \right],~~{\text{up~to}~\cO(\st^{-1})},~~\text{where}, \label{P-Q-PL} \\
 \cM^{PL,\cQ}_\rhoi(Y,\theta) &=& \frac{1 }{A^{\frac{1}{\alpha}} \alpha}  \int_{0}^\infty \dd Z~ Z^{\frac{1}{\alpha}-1} \left[\log \left( 1+ (\ee^{\ci \theta}-1) G_+(Y+Z) \right) + \log \left( 1+ (\ee^{-\ci \theta}-1) G_-(Y-Z) \right)  \right].~~~~~ \label{cM-Q-PL}
\end{eqnarray}
This exact expression of the distribution of the displacement of the tagged particle for the power law initial densities is compared with numerical simulation in fig. \ref{fig:varCIC}a. 
From this distribution one can obtain the scaling of various moments with respect to time (through $\st$) as follows: Assuming that the $\theta$ integral may be performed by the saddle-point method, up to subexponential factors we have $P^\cQ_\rhoi(y,t) \sim \exp[\st^\frac{1}{\alpha} \cM^{PL, \cQ}_\rhoi (y/\st,\thetast)]$ for some saddle point $\thetast$. The cumulants of $y$ are given by the generating function
\begin{eqnarray}
 \log \langle \ee^{\lambda y} \rangle &\sim& \log \left( \int \dd y \exp \left[ \lambda y + \st^\frac{1}{\alpha} \cM^{PL,\cQ}_\rhoi (Y,\thetast) \right] \right)= \log \left( \int \dd y  \exp \left[ \st^\frac{1}{\alpha} \left(\lambdat Y +  \cM^{PL,\cQ}_\rhoi (Y,\thetast) \right) \right]  \right) = \st^\frac{1}{\alpha} \mu(\lambdat), ~~~~~~
\end{eqnarray}
with $\lambdat = \st^{1-\frac{1}{\alpha}} \lambda$.
In the last equality $\mu$ is the cumulant generating function, which is also the Legendre transform of $\cM^{PL, \cQ}_\rhoi (Y,\theta)$. The cumulants $\kappa_n$ are obtained by taking derivatives of $\mu$,
\begin{equation}
 \label{eq:cumy}
 \kappa_n \propto \frac{\partial^n (\st^\frac{1}{\alpha} \mu(\lambdat))}{\partial \lambda^n} = \st^\frac{1}{\alpha} (\st^{1-\frac{1}{\alpha}})^n \frac{\partial^n \mu(\lambdat)}{\partial \lambdat^n} \propto \st^{n-\frac{n-1}{\alpha}}.
\end{equation}
For larger and larger $\alpha$ one gets closer and closer to the free particle, as the surroundings are less and less crowded, while for small $\alpha$ the cumulants are very small. The uniform case $\alpha=1$ is intermediate. The prediction on the scaling with respect to $t$ is verified in Fig.\,\ref{fig:varCIC}b where we plot the variance of the displacement of a brownian TP (moving inside the pool of other hardcore Brownian particles) for power law quenched initial position configurations with $\alpha=2$ from simulation. We observe that the variance grows as $\sim t^{3/4}$ as expected because $\st=\sqrt{2t}$ for Brownian propagators.

\subsection{Annealed case}

In the annealed case the analysis is similar to step annealed case. From the expression for the density in \eqref{eq:rhoipow}, it is clear that $A(y_0,t)$ and $B(y_0,t)$ are proportional to $\st^{1/\alpha}$. Hence the large deviation function also has the same scaling $\log P^{PL,\cA}_{\rhoi} (y_0,t) \sim -\st^{1/\alpha} \cF^{PL,\cA}_\rhoi(Y)$ with respect to $\st$. 
Inserting \eqref{eq:rhoipow} in \eqref{F-A-any}, we get the following explicit form of the large deviation function
\begin{equation}
 \label{F-A-PL}
 \cF^{PL,\cA}_\rhoi(Y) = \frac{1}{A^{1/\alpha} \alpha} \left[\sqrt{\int_{Z=-\infty}^0 \dd Z |Z|^{1/\alpha-1} G_+(Y-Z)} - \sqrt{\int_{Z=0}^\infty \dd Z |Z|^{1/\alpha-1} G_-(Y-Z)}  \right]^2.
\end{equation}
Since the scaling of the large deviation function is the same as in the quenched case, we therefore expect the cumulants to have the same scaling as well, $\kappa_n \propto \st^{n-\frac{n-1}{\alpha}}$.

\section{Exact Joint distribution of displacements of two tagged particles}
\label{section:2TPs}
Until now we have looked at the position distribution of a single tagged particle in a single file motion. In this section we derive the joint distribution of the positions of two tagged particles. More precisely, we provide an explicit exact expression of the joint propagator $P_2(y_0,y_\ell ,t | \bx,0)$ of the positions $y_0$ and $\yl$ of, respectively, the initial $0$th and the $\ell$th particles, given the initial positions of all the particles 
$-\infty < x_{-N}\le x_{-N+1}\le...\le x_0 \le...\le x_{N-1} \le x_{N}< \infty $. The procedure for the single TP as done  in sec. \ref{section:galcalc}, can be easily extended for two TP case as follows: Since the particles are identical and hard-core interacting, one can use the non-interacting mapping \cite{Sadhu-15}. In this equivalent non-interacting picture, starting from the initial positions $\bx=(x_{-N}, x_{-N+1},...,x_0,...,x_{N-1}, x_{N})$, any two particle say $n$th and $m$th particle reach $y_0$ and $\yl$ and rest of the $2N-1$ particles go in one of the flowing three regions:  (I) $(-\infty, y_0)$, (II) $(y_0, \yl)$ and (III) $(\yl,\infty)$ in such a way that the number of particles in regions (I), (II) and (III) are, respectively, $N$, $\ell-1$ and $N-\ell$. Here, without any loss of generality, we have assumed $\ell >0$. 
As a result the particle at $y_0$ and $\yl$ becomes, respectively,  the $0th$ and $\ell$th particle at time $t$. Hence the propagator can be written as : 
\begin{eqnarray}
\label{eq:P2yt1}
%P_2(y_0,\yl,t|\bx) &=& \sum_{m=-N}^N \sum_{\stackrel{n=-N}{n\neq m}}^Ng(y_0,t|x_{m},0) g(\yl,t|x_{n},0)  \left(\prod_{\stackrel{k=-N}{{k\neq m}, {k \neq n}}}^N\sum_{\epsilon_{k}=(-1,0,1)}\right) \nonumber \\
%&&~~~~~~~~
%\times~~ \delta_{(\sum_{\stackrel{k=-N}{ k\neq m, k \neq n}}^{N}\epsilon_k),\ell} ~ \times~\delta_{(\sum_{\stackrel{k=-N}{ k\neq m, k \neq n}}^{N}\epsilon_k^2),2N-\ell}~ 
%~ \prod_{\stackrel{k=-N}{{k\neq m}, {k \neq n}}}^N g_{\epsilon_{k}}(y_0,\yl,x_k,t), ~~\text{where}, \\
%g_1(y_0,\yl, x_k, t) &=& \int_{-\infty}^{y_0} dy~g(y,t|x_k,0), 
%~~g_0(y_0,\yl, x_k, t)= \int_{y_0}^{\yl} dy~g(y,t|x_k,0),
%~~g_{-1}(y_0,\yl, x_k, t)= \int_{\yl}^{\infty} dy~g(y,t|x_k,0). \nonumber 
P_2(y_0,\yl,t|\bx) &=& \sum_{m=-N}^N \sum_{\stackrel{n=-N}{n\neq m}}^Ng(y_0,t|x_{m},0) g(\yl,t|x_{n},0)  \sum_{\epsilon_{-N}=(-1,0,1)}\sum_{\epsilon_{-N+1}=(-1,0,1)}... \sum_{\epsilon_{N-1}=(-1,0,1)}\sum_{\epsilon_{N}=(-1,0,1)} \nonumber \\
&&~~~~~~~~
\times~~ \delta_{(\sum_{\stackrel{k=-N}{ k\neq m, k \neq n}}^{N}\epsilon_k),\ell} ~ \times~\delta_{(\sum_{\stackrel{k=-N}{ k\neq m, k \neq n}}^{N}\epsilon_k^2),2N-\ell}~ 
~ \prod_{\stackrel{k=-N}{{k\neq m}, {k \neq n}}}^N g_{\epsilon_{k}}(y_0,\yl,x_k,t), ~~\text{where}, \\
g_1(y_0,\yl, x_k, t) &=& \int_{-\infty}^{y_0} dy~g(y,t|x_k,0), 
~~g_0(y_0,\yl, x_k, t)= \int_{y_0}^{\yl} dy~g(y,t|x_k,0),
~~g_{-1}(y_0,\yl, x_k, t)= \int_{\yl}^{\infty} dy~g(y,t|x_k,0). \nonumber 
\end{eqnarray}
Replacing the Kronecker deltas by their integral representations, we get for $\yl > y_0$,
\begin{eqnarray}
\label{eq:P2yt1}
P_2(y_0,\yl,t|\bx) &=&  \frac{1}{4 \pi^2} \int_{-\pi}^{\pi} \dd \theta \int_{-\pi}^{\pi} \dd \phi~\ee^{\ci \theta \ell} 
\ee^{\ci \phi (2N-\ell)}~\times~\sum_{m=-N}^N \sum_{\stackrel{n=-N}{n\neq m}}^Ng(y_0,t|x_{m},0) g(\yl,t|x_{n},0)  \nonumber \\
&& ~~~~~~~~~~~~~~~~~~~ 
\times~\left(\prod_{\stackrel{k=-N}{{k\neq m}, {k \neq n}}}^N\sum_{\epsilon_{k}=(-1,0,1)}
\ee^{-\ci (\epsilon_k\theta +\epsilon_k^2 \phi)} ~g_{\epsilon_{k}}(y_0,\yl,x_k,t) \right), \nonumber \\
&=& \frac{1}{4 \pi^2} \int_{-\pi}^{\pi} \dd \theta \int_{-\pi}^{\pi} \dd \phi~\ee^{\ci \theta \ell} 
\ee^{\ci \phi (2N-\ell)}~\times~\sum_{m=-N}^N \sum_{\stackrel{n=-N}{n\neq m}}^Ng(y_0,t|x_{m},0) g(\yl,t|x_{n},0)  \nonumber \\
&& ~~~~~
\times~\prod_{\stackrel{k=-N}{{k\neq m}, {k \neq n}}}^N\left( \ee^{-\ci(\phi-\theta)} g_{1}(y_0,\yl,x_k,t) + g_{0}(y_0,\yl,x_k,t)+
\ee^{-\ci(\phi+\theta)} g_{-1}(y_0,\yl,x_k,t)\right), \nonumber 
%&=& \frac{\partial^2}{\partial y_0 \partial \yl} \left[ -\frac{1}{16 \pi^2} \int_{-\pi}^{\pi} \dd \theta \int_{-\pi}^{\pi} \dd \phi~\frac{\ee^{-\ci \theta \ell} \ee^{-\ci \theta (2N-\ell)}}{\sin(\theta-\phi)\sin(\theta+\phi)} \right. \nonumber \\ 
%&& \left. \times~\prod_{\stackrel{k=-N}{{k\neq m}, {k \neq n}}}^N\left( \ee^{-\ci(\phi-\theta)} g_{1}(y_0,\yl,x_k,t) + g_{0}(y_0,\yl,x_k,t)+
%\ee^{-\ci(\phi+\theta)} g_{-1}(y_0,\yl,x_k,t)\right)\right] 
\end{eqnarray}
After some manipulations we finally get the following exact expression for $\yl \ge y_0$:
\begin{eqnarray}
\label{eq:2part}
P_2(y_0,\yl,t|\bx) &=& \frac{\partial^2}{\partial y_0 \partial \yl} \left[ \frac{1}{16 \pi^2} \int_{-\pi}^{\pi} \dd \theta \int_{-\pi}^{\pi} \dd \phi~\frac{\ee^{\ci (\theta -\phi)\ell} \ee^{\ci (2N+1)\phi }}{\sin[(\phi-\theta)/2]\sin[(\theta+\phi)/2]}~~D_N(y_0,\yl,\theta,\phi,t|\bx) \right], ~~\text{where}, \nonumber \\ 
D_N(y_0,\yl,\theta,\phi,t|\bx)&=& \prod_{k=-N}^N\left( \ee^{-\ci(\phi-\theta)} g_{1}(y_0,\yl,x_k,t) + g_{0}(y_0,\yl,x_k,t)+\ee^{-\ci(\phi+\theta)} g_{-1}(y_0,\yl,x_k,t)\right).
\end{eqnarray}
This is our fourth main result. This expression provides the exact joint propagator of $0$th and $\ell$th ($\ell >0$) tagged particles whose normalisation property can easily be checked using the identity 
\begin{eqnarray}
\frac{1}{2 \pi} \int_{-\pi}^\pi  \dd \psi ~\frac{ \sin \left[ (2m+1)\psi \right]}{\sin  \psi } = 1, ~~\forall~m \in \mathbb{Z}_+,\label{norm-chk-2}
\end{eqnarray}
similar to what is done in sec. \ref{section:galcalc}. The $l< 0$ case can be obtained analogously. From \eqref{eq:2part} one can easily obtain the thermodynamic limit as well as the large deviation functions similar to the single TP case in sec. \ref{Thermodynamic-limit}. 

%\newpage

\section{Conclusion}
\label{section:ccl}

In this work we have presented very general calculations of the propagator of a tagged particle in single-file system for arbitrary initial position configurations. This leads us to widely applicable exact expressions for the propagator in the thermodynamic limit for a particular class of single particle propagators with arbitrary initial density profiles both in the quenched and the annealed settings. In the quenched case the propagator is expressed in terms of an finite integral over $\theta$,  while in the annealed systems it can be expressed as a very simple combination of Bessel functions. In particular we have found that, in the annealed case the large deviation function for arbitrary initial density profile is a simple generalisation of the one obtained for flat/uniform initial density profile~\cite{Hegde-14, Krapivsky-15, Sadhu-15}, whereas in the quenched case it is expressed implicitly. We particularly have focused on the step and the power law initial density profiles.  We have shown that an average drift is generated in the step case, and  in the power-law case the scaling with respect to time changes from what one observes in the flat case. 

The procedure presented here is simpler than earlier hydrodynamic and microscopic methods \cite{Krapivsky-14, Krapivsky-15, Sadhu-15, Hegde-14, Sabhapandit-15}. This method relies on the following three important facts: (i) all the particles are identical, (ii) it is possible to describe the system on the single particle level through some well defined propagator and (iii) the inter-particle interactions are hardcore repulsive so that one can use the mapping to non-interacting particles. Moreover our method can be easily generalised to the case of multiple tagged particles as shown in sec. \ref{section:2TPs} for two tagged particles. We also claim that it can be extended to compute joint unequal time distributions as well as various multi-point correlations and cumulants exactly. 

There are several other directions in which our work can be extended. For example, how to compute the TP displacement distribution in cases where there is no single-particle description, \textit{e.g.} in Random Average Process ? It would also be interesting to compute this distribution in situations where inter-particle interactions are more complicated than hardcore repulsion, for example, symmetric simple exclusion process. A more ambitious challenge would be to obtain such exact results in the presence of a locally driven tracer.

\vspace{0.25cm}
We thank Tridib Sadhu and David Mukamel for helpful discussions.

\end{document}